%% file: BAM493_pubcom.tex
\documentclass[prl,twocolumn,showpacs,amsmath,amssymb]{revtex4-1}
\usepackage{subfigure}
\usepackage{graphicx}
\usepackage{graphicx}
\usepackage{array}
\usepackage{xcolor}

\usepackage{lineno}
\usepackage{LamcNPi-defs}

\begin{document}
\normalsize
\parskip=5pt plus 1pt minus 1pt

\title{Observation of the Singly Cabibbo-Suppressed Decay \LamCNPi{}}

\input{author.tex}

\begin{abstract}

The singly Cabibbo-suppressed decay \LamCNPi{} is observed for the first time with a statistical significance of $7.3\sigma$ by using 3.9~\ifb{} of \ee{} collision data collected at  center-of-mass~energies between 4.612 and 4.699~GeV with the BESIII detector at BEPCII.
The branching fraction of \LamCNPi{} is measured to be $(6.6\pm1.2_{\rm stat}\pm0.4_{\rm syst})\times 10^{-4}$. 
By taking the upper limit of branching fractions of $\LamC\to p\pi^0$ from the Belle experiment, the ratio of branching fractions between \LamCNPi{} and $\LamC\to p\pi^0$ is calculated to be larger than 7.2 at the 90\% confidence level, which disagrees with the most predictions of available phenomenological models.
In addition, the branching fractions of the Cabibbo-favored decays \LamCLamPi{} and \LamCSigPi{} are measured to be $(1.31\pm0.08_{\rm stat}\pm0.05_{\rm syst})\times 10^{-2}$ and  $(1.22\pm0.08_{\rm stat}\pm0.07_{\rm syst})\times 10^{-2}$, respectively, which are consistent with previous results.

\end{abstract}

\maketitle

The decay of the ground state charmed baryon \LamC{} plays an essential role in studying the nature of both strong and weak interactions in heavy-to-light baryonic transitions~\cite{Cheng:2015iom}. 
The hadronic decay amplitudes of \LamC{} consist of factorizable and nonfactorizable components, in which the nonfactorizable effects arising from $W$ exchange and internal $W$ emission play an essential role~\cite{Yang:1986SU3, Kohara:1991}. Therefore, studies of nonfactorizable components are critical to understanding the underlining dynamics of charmed baryon decays. 

In the last three decades, several different phenomenological models, e.g., current algebra~\cite{Cheng:1993gf, Sharma:1998rd} and SU(3) flavor symmetry~\cite{Sharma:1996sc, Lu:2016ogy,  Geng:2017esc}, have been employed as tools to reveal the dynamics of charmed baryon decays. The nonfactorizable contributions are important in these decays, in contrast to the meson case, and can be constrained by measurements.
Studies of singly Cabibbo-suppressed (SCS) decay modes containing both factorizable and nonfactorizable contributions will provide information about their interference; therefore comprehensive and precise experimental inputs are required for an improved understanding of the validity of the different phenomenological models.
Experimentally, great progress has been made in the study of charmed-baryon decays recently, particularly for Cabibbo-favored (CF) decays~\cite{Belle:2017tfw, Ablikim:2018czr, Ablikim:2020ffk, Li:2021iwf}; for example,
the branching fraction of the golden mode $\LamC\to pK^{-}\pi^+$ and the neutron final-state decay $\LamC\to n\Ks \pi^+$ have been measured with a precision of better than 10\%~\cite{Zupanc:2013iki, Ablikim:2015flg, Ablikim:2016mcr}. 
However, experimental studies of the SCS decays are still quite challenging due to their small branching fractions of $10^{-3}$ or below.

The two-body SCS decay \LamCNPi{}, together with  $\LamC\to p\pi^0$ and $\LamC\to p\eta$, are of great interest and have been studied extensively in the context  of various phenomenological models~\cite{Uppal:1994pt, Sharma:1996sc, Chen:2002jr, Cheng:2018hwl, Lu:2016ogy, Geng:2018plk, Geng:2018rse, Geng:2020zgr, Zou:2019kzq, HJZhao:2020}. 
Compared to $\LamC\to p\eta$, the decays \LamCNPi{} and $\LamC\to p\pi^0$ are suppressed due to the destructive interference between the factorizable and nonfactorizable amplitudes~\cite{Cheng:2018hwl}. 
Different phenomenological models predict quite different decay rates for  $\LamC\to p\pi^0$ and \LamCNPi{}, and distinguishing between these models with experimental results is highly desirable. 
The ratio of the  branching fractions between \LamCNPi{} and $\LamC\to p\pi^0$ is a particularly sensitive observable in comparing these models, since correlated uncertainties in theoretical calculation can be canceled. 
This ratio is predicted to be 2 by the SU(3) flavor symmetry model~\cite{Sharma:1996sc,  Lu:2016ogy, Geng:2018plk}, 4.5 or 8.0 by the constituent quark model~\cite{Uppal:1994pt}, 3.5 by a dynamical calculation based on pole model and current-algebra~\cite{Cheng:2018hwl}, 4.7 by the SU(3) flavor symmetry including the contributions from $\mathcal{O}(\overline{15})$~\cite{Geng:2018rse}, and 9.6 by the topological-diagram approach~\cite{HJZhao:2020}.

To date, the branching fractions of only a few SCS decay modes have been measured, and all  with limited precision~\cite{PDG:2020}, and those involving a neutron in the final state have never been measured. BESIII and Belle have reported the branching fractions of $\LamC\to p\eta$ and the upper limit for $\LamC\to p\pi^0$~\cite{Ablikim:2017ors, Li:2021uzo}, where  for $\LamC \to p\pi^{0}$ some tension exists between measurement and  some of the predictions~\cite{Uppal:1994pt, Sharma:1996sc,  Lu:2016ogy, Geng:2018plk}.
Therefore, an observation of the decay  \LamCNPi{} is essential for validating and constraining different dynamical models.

In this Letter, the first observation of the SCS decay \LamCNPi{} is reported using 3.9~\ifb{} \ee{} collision data collected with the BESIII detector at six center-of-mass (c.m.)~energies between 4.612 and 4.699~GeV. 
The integrated luminosities of the data samples at 4.612, 4.628, 4.641, 4.661, 4.682, and 4.699~GeV 
are 103.5, 519.9, 548.2, 527.6, 1664.3, and 534.4~pb$^{-1}$~\cite{BESIII:Lumi}, respectively. 
Throughout this Letter, charge-conjugate modes are implicitly included. 

A detailed description of the design and performance of the BESIII detector can be found in Ref.~\cite{Ablikim:2009aa}.
Simulated samples are produced with a GEANT4-based~\cite{Agostinelli:2002hh} Monte Carlo (MC) package, which includes the geometric description of the BESIII detector. The signal MC samples of $\ee\to\LamC\ALamC$ with \ALamC{} decaying into ten specific tag modes
 (as described below and listed in Table~\ref{tab:yield-st-468}) and \LamCNPi, $\Lambda\pi^+$,  and $\Sigma^0\pi^+$ are generated for each individual c.m.~energy by the generator {\sc kkmc}~\cite{Jadach:2000ir} 
by incorporating initial-state radiation (ISR) effects and the beam energy spread. 
The inclusive MC sample, which consists of \LCLC{} events, $D_{(s)}$ production, ISR return to lower-mass $\psi$ states, and continuum processes $e^{+}e^{-}\rightarrow q\bar{q}$ ($q=u,d,s$), is generated to estimate the potential background, in which all the known decay modes of charmed hadrons and charmonia are modeled with {\sc evtgen}~\cite{Lange:2001uf, Ping:2008zz} using branching fractions taken from the Particle Data Group~\cite{PDG:2020}, and the remaining unknown decays are modeled with {\sc lundcharm}~\cite{Chen:2000tv}. 
Final-state radiation from charged final-state particles is incorporated using  
{\sc photos}~\cite{Richter-Was:1992hxq}.

A double-tag (DT) approach~\cite{MarkIII:DT} is implemented to search for \LamCNPi{}. A data sample of \ALamC{} baryons, referred to as the single-tag (ST) sample, is reconstructed with ten exclusive hadronic decay modes, as listed in Table~\ref{tab:yield-st-468}, where the intermediate particles \Ks, $\bar{\Lambda}$, $\bar{\Sigma}^0$, $\bar{\Sigma}^-$, and $\pi^0$ are reconstructed with the decays 
$\Ks\to \pi^+\pi^-$, $\bar{\Lambda}\to \bar{p}\pi^+$, $\bar{\Sigma}^0\to\gamma\bar{\Lambda}$, $\bar{\Sigma}^-\to \bar{p}\pi^0$, and $\pi^0\to \gamma\gamma$, respectively.
Those events in which the signal decay \LamCNPi{} is reconstructed in the system recoiling against the \ALamC{} candidates of the ST sample are denoted as DT candidates. 

Charged tracks detected in the helium-based main drift chamber (MDC) are required to be within a polar angle ($\theta$) range of $|\!\cos\theta| < 0.93$, where $\theta$ is defined with respect to the beam direction.
Except for those from \Ks{} and $\Lambda$ decays, their distances of the closest approach to the interaction point (IP) are required to be within $\pm$10~cm along the beam direction and 1~cm in the plane perpendicular to the beam (referred to as tight track hereafter).
The particle identification (PID) is implemented by combining measurements of the energy deposited in the MDC ($dE/dx$) and the flight time in the time-of-flight system, and every charged track is assigned a particle type of pion, kaon or proton, according to which  assignment has  the highest probability. 

Photon candidates are identified using showers in the electromagnetic calorimeter (EMC). The deposited energy of each shower must be more than 25~MeV in the barrel region ($|\!\cos\theta| \le 0.80$) or more than 50~MeV in the end-cap region ($0.86 \le |\!\cos\theta| \le 0.92$). To suppress electronic noise and showers unrelated to the event, the difference between the EMC time and the event start time is required to be within (0, 700)~ns. The $\pi^0$ candidate is reconstructed with a photon pair within the invariant-mass region (0.115, 0.150)~GeV/$c^2$. To improve the resolution, a kinematic fit is performed by constraining the invariant mass of the photon pair to be the $\pi^0$ mass and requiring the corresponding $\chi^2$ of the fit to be less than 200.
The momenta updated by the kinematic fit are used in the further analysis.

Candidates for \Ks{} and $\bar{\Lambda}$ mesons are reconstructed in their decays to $\pi^+\pi^-$ and $\bar{p}\pi^+$, respectively, where the charged tracks must have distances of closest approaches to the IP that are within $\pm$20~cm along the beam direction (referred to as loose track hereafter).
To improve the signal purity, PID is applied to the (anti)proton candidate, while the charged pion is not subjected to a PID requirement. 
A secondary vertex fit is performed to each \Ks{} or $\bar{\Lambda}$ candidate, and the momenta updated by the fit are used in the further analysis. A \Ks{} or $\bar{\Lambda}$ candidate is accepted by requiring the $\chi^2$ of the secondary vertex fit to be less than 100. 
Furthermore, the decay vertex is required to be separated from the IP by a distance of at least twice the fitted vertex resolution, and the invariant mass to be within (0.487, 0.511)~GeV/$c^2$ for $\pi^+\pi^-$  or (1.111, 1.121)~GeV/$c^2$ for $\bar{p}\pi^+$.
The $\bar{\Sigma}^0$ and $\bar{\Sigma}^-$ candidates are reconstructed with the $\gamma\bar{\Lambda}$
and $\bar{p}\pi^0$ final states, requiring the invariant masses to lie within (1.179, 1.203) and (1.176, 1.200)~GeV/$c^2$, respectively.

The ST \ALamC{} candidates are identified using the variables of beam-constrained invariant mass $M_\mathrm{BC} = \sqrt{\Ebeam^2/c^4 - | \pALC |^2/c^2}$ and energy difference $\dE = \Ebeam - E_{\ALamC}$, where \Ebeam{} is the beam energy, $E_{\ALamC}$ and \pALC{} are the energy and momentum of the \ALamC{} candidate, respectively. 
The \ALamC{} candidate is required to satisfy tag-mode dependent \dE{} requirements, the asymmetric intervals of which take into account the effects of ISR and correspond to three times the resolution around the peak, as summarized in Table~\ref{tab:yield-st-468}. If there is more than one candidate satisfying the above requirements for a specific tag mode, the one with the smallest $|\dE|$ is kept. 

\begin{table}[!htbp]
  \begin{center}
  \caption{\dE{} requirement, the ST yield, and the detection efficiency of the ST and DT $\LamCNPi$ selections for  
            each tag mode of the  data sample at  $\sqrt{s}=4.682$~GeV. The uncertainty on the ST yield is statistical only. }
  \renewcommand\arraystretch{1.2}
    \begin{tabular}{ l c p{1cm}<{\raggedleft} @{ $\pm$ } p{1cm}<{\raggedright} c c}
      \hline
      \hline
	    & \dE{} (MeV) & \multicolumn{2}{c}{$N_{i}^{\mathrm{ST}}$}  & $\epsilon_{i}^{\mathrm{ST}}$(\%) & $\epsilon_{i}^{\mathrm{DT}}$ (\%) \\
      \hline
	                    
            $\apkpi$            & $(-34,~20)$    &  $17,415$  & 145  & 47.3  &  37.0 \\
            $\apks$             &  $(-20,~20)$   &  $3,353$ & 61  &  48.1  & 38.8 \\
            $\apkpi\pi^0$     &  $(-30,~20)$        &  $4,005$ & 95  & 14.5  & 13.4 \\
            $\apks\pi^0$          & $(-30,~20)$  &  $1,454$ & 52    &  16.5  & 14.4 \\
            $\apks\pi^+\pi^-$    & $(-20,~20)$    &  $1,261$ & 49   &  17.7  & 14.8 \\
            $\bar{\Lambda}\pi^-$      &  $(-20,~20)$     &  $2,012$ & 47   &  37.8  & 31.0 \\
            $\bar{\Lambda}\pi^-\pi^0$    & $(-30,~20)$ &  $3,576$ & 71  &  14.6  & 12.9 \\
            $\bar{\Lambda}\pi^-\pi^+\pi^-$ & $(-20,~20)$ &  $1,818$ & 52   &  12.3  & 10.3 \\
            $\bar{\Sigma}^0\pi^-$        & $(-20,~20)$  &  $1,047$ & 34   &  19.3  & 17.4 \\
            $\bar{\Sigma}^-\pi^+\pi^-$   & $(-30,~20)$ &  $2,275$ &  63  &  16.2  & 16.1 \\
      \hline\hline
    \end{tabular}
          \label{tab:yield-st-468}
  \end{center}
\end{table}

For the $\ALamC\to\apks\pi^0$ ST mode, candidate events with $M_{\bar p\pi^+} \in (1.110,1.125)$~GeV/$c^2$ and $M_{\bar p\pi^0}\in (1.170,1.200)$~GeV/$c^2$ are vetoed to avoid double counting with the $\ALamC\to\bar{\Lambda}\pi^-\pi^0$ or $\ALamC\to\bar{\Sigma}^-\pi^+\pi^-$ ST modes, respectively. For the $\ALamC\to\bar{\Sigma}^-\pi^+\pi^-$ ST mode, candidate events with $M_{\pi^+\pi^-}\in (0.490,0.510)$~GeV/$c^2$ and $M_{\bar p\pi^+}\in (1.110,1.125)$~GeV/$c^2$  are rejected to avoid double counting with the $\ALamC\to\apks\pi^0$ or $\ALamC\to\bar{\Lambda}\pi^-\pi^0$ ST modes, respectively.
In the $\ALamC\to\apks\pi^+\pi^-$ and $\bar{\Lambda}\pi^-\pi^+\pi^-$ selections, candidate events with $M_{\bar p\pi^+}\in (1.110,1.125)$~GeV/$c^2$ and $M_{\pi^+\pi^-}\in (0.490,0.510)$~GeV/$c^2$ are rejected, respectively. 

The \mBC{} distributions of surviving candidates for the ten ST modes  are illustrated in Fig.~\ref{fig:single-tag-468} for the data sample at $\sqrt{s}=4.682~\mathrm{GeV}$, where clear \ALamC{} signals are observed in each sample. No peaking backgrounds are found with the investigation of the inclusive MC sample. To obtain the ST yields, unbinned maximum likelihood fits on these $M_{\rm BC}$ distributions are performed, where the signal shapes are modeled with the MC-simulated shape convolved with a Gaussian function representing the resolution difference between data and MC simulation, and the background shapes are described by an ARGUS function~\cite{ARGUS:1990hfq}. 
The candidates with $M_{\rm BC} \in (2.275,2.31)$~GeV/$c^2$ are retained for further analysis, and the signal yields for the individual ST modes are summarized in Table~\ref{tab:yield-st-468}. The same procedure is performed for the other five data samples at different c.m.~energies. The ST yields of the other five samples with different c.m. energies are summarized in the Supplemental Material~\cite{BESIII:Supplemental}. The sum of the ST yields for all the six data samples is $90,692\pm359$, where the uncertainty is statistical. 

\begin{figure}[!htp]
    \begin{center}
        \includegraphics[width=0.35\textwidth, trim=5 0 0 0, clip]{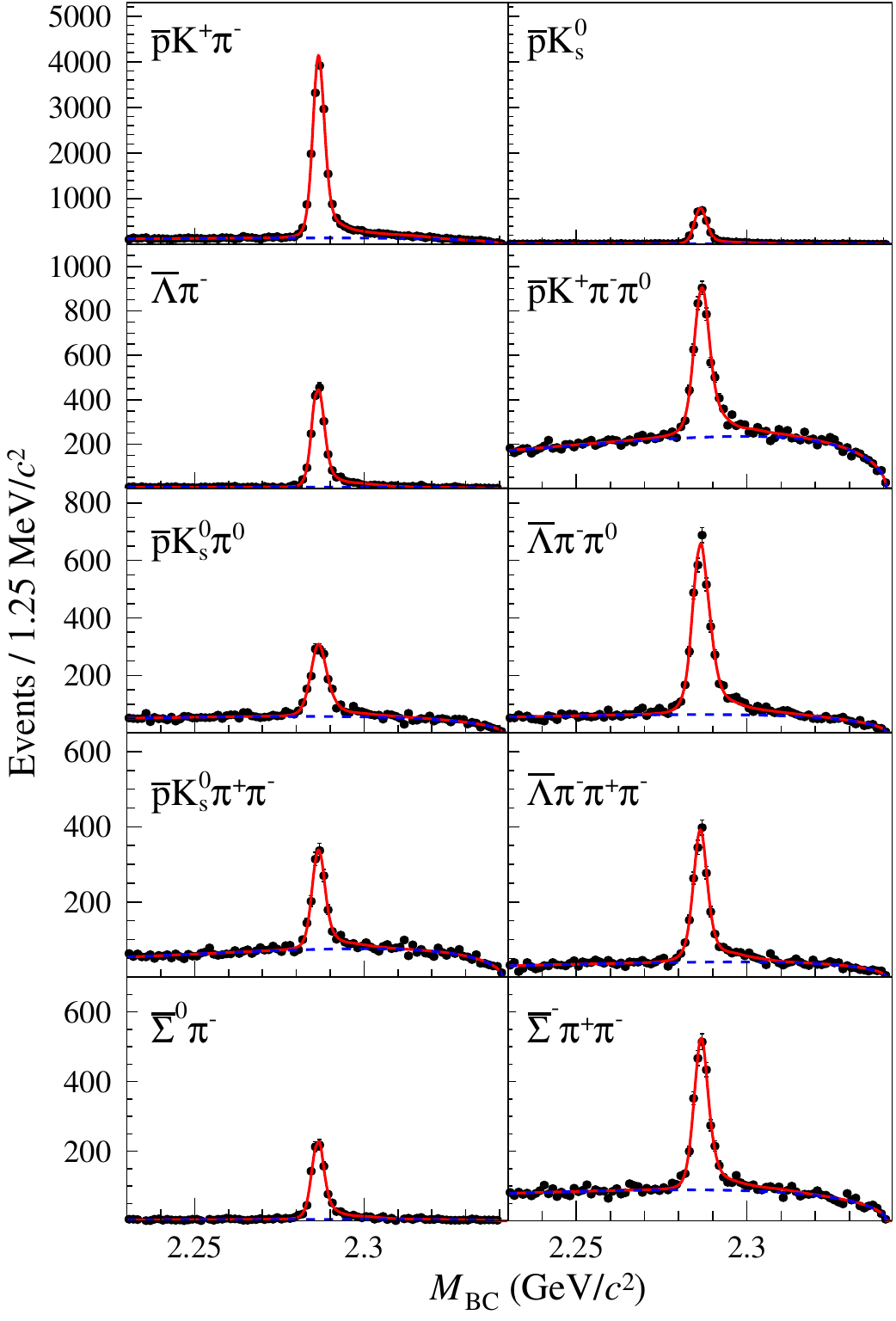}
    \end{center}
    \caption{
      The \mBC{} distributions of the ST modes for data sample at 
      $\sqrt{s}=4.682~\mathrm{GeV}$. 
      The points with error bars represent data. The (red) solid curves indicate the fit results and the (blue) dashed curves describe
      the background shapes.}
     \label{fig:single-tag-468}
\end{figure}

The decay \LamCNPi{}  is searched for among the remaining tracks recoiling against the ST \ALamC{} candidates.
Only one tight charged track is allowed, which is then assigned to be the $\pi^+$ from the signal decay. 
To suppress contamination from  long-lifetime particles in the final state, the candidate events are further required to be without any loose tracks.
It reduces the signal efficiency of \LamCNPi{} by 6\% and background level by 50\%. Meanwhile, the signal yields for \LamCLamPi{} and $\Sigma^0\pi^+$ will decrease significantly as the $\Lambda$ and $\Sigma^{0}$ decays with charged particles in the final state are highly suppressed, and their decays with neutral ones mostly pass this requirement.
To improve detection efficiency, the neutron is selected through  the recoiling mass ($M_{\rm rec}$) against  the ST \ALamC{} and $\pi^+$:
\begin{equation} \label{eq:mrec}
  M^2_{\rm rec} = (\Ebeam - E_{\pi^+})^2/c^4 - | \rho\cdot\vec{p}_{0} - \vec{p}_{\pi^+} |^2/c^2 ,
\end{equation}
where $E_{\pi^+}$ and $\vec{p}_{\pi^+}$ are the energy and momentum of the $\pi^+$ candidate,
$\rho = \sqrt{\Ebeam^2/c^2 - m_{\LamC}^2 c^2}$, and $\vec{p}_{0} = - \pALC/|\pALC|$ is the unit direction opposite to the 
ST \ALamC{}. 

After imposing all selection conditions mentioned above, the distribution of  $M_{\rm rec}$ of the accepted DT candidate events from the combined six data samples at different c.m.~energies is shown in Fig.~\ref{fig:fit}, where a peak at the neutron mass is observed, representing the  \LamCNPi{} signal.
Additionally, there are two prominent structures peaking at the $\Lambda$ and $\Sigma^0$ mass regions, 
which represent the CF processes  \LamCLamPi{} and \LamCSigPi{}, respectively.

\begin{figure}[!htp]
    \begin{center}
            \includegraphics[width=0.35\textwidth,height=0.19\textheight, trim=5 0 0 0, clip]{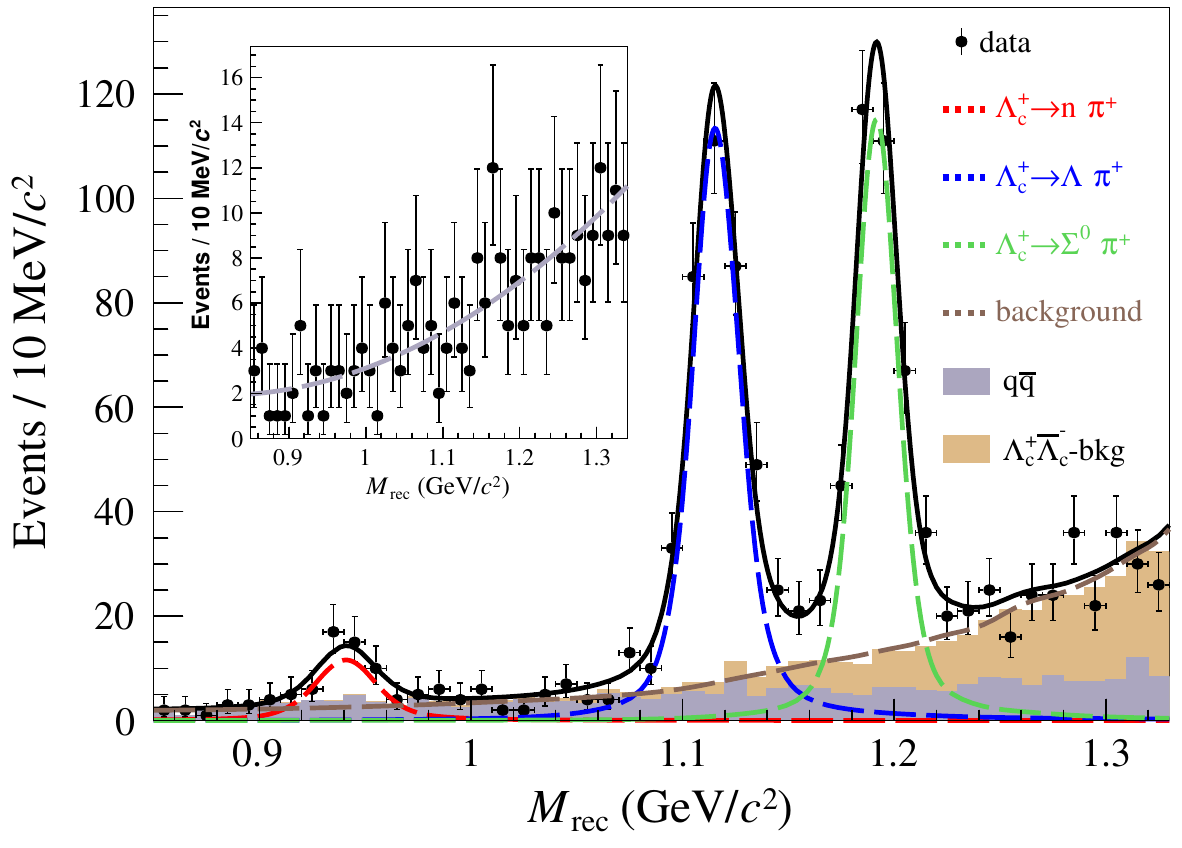}   
	\end{center}
    \caption{ The $M_{\mathrm{rec}}$ distribution of the accepted DT candidate events from the combined six data samples.
                  The black points with error bars are data. The red, blue, and green dashed lines indicate the curves for the neutron, 
                  $\Lambda$, and $\Sigma$ peaks, respectively. 
                  The brown and gray shaded histograms for the two background components are from the inclusive MC sample, and the dark brown dashed line indicates the curve that describes the two background components from fitting.
                  The black line is the sum over all the components in the fit. The inset shows the $M_{\mathrm{rec}}$ distribution in the ST \mBC{} sideband region, and the gray dashed line indicates the curve that describes the $q\bar{q}$ background.}
        \label{fig:fit}
\end{figure}

The potential backgrounds can be classified into two categories: those directly originating from continuum hadron production in the \ee{} annihilation (referred to as $q\bar{q}$ background hereafter) and those from \ee{}$\to$\LamC{}\ALamC{} events (referred to as  \LamC{}\ALamC{} background hereafter), excluding contributions from \LamCNPi, $\Lambda\pi^+$, and $\Sigma^0\pi^+$ signals.
The distributions and magnitudes of $q\bar{q}$ and \LamC{}\ALamC{} backgrounds are estimated with the inclusive MC sample, as shown in Fig.~\ref{fig:fit}, where no peaking backgrounds are observed.
The understanding of the $q\bar{q}$ background is also validated with candidate events in the $\mBC$ sideband region $(2.235,2.270)$~GeV/$c^2$ of ST candidates in data; here also,  no peaking background is observed.

The signal yield $N_{\rm obs}$ is obtained by performing an unbinned maximum-likelihood fit on the 
 $M_{\rm rec}$ distribution. The  neutron, $\Lambda$ and $\Sigma^0$ signals are modeled by the MC-simulated shapes convolved with Gaussian functions that account for the resolution difference between data and MC simulation, and where the three Gaussian functions share the same width parameters.
The $q\bar{q}$ background is described by a second-order Chebyshev function with fixed parameters, 
which are obtained by fitting the corresponding distribution of events in the ST \mBC{} sideband region.
The  shape of the  \LCLC{} background is taken from the inclusive MC sample.
The fit distributions are depicted in Fig.~\ref{fig:fit}, and correspond to signal yields of $50 \pm 9$,  $376 \pm 22$, and $343 \pm 22$ for the decays \LamCNPi{}, $\Lambda \pi^+$, and $\Sigma^0 \pi^+$, respectively, where the uncertainties are statistical.
The statistical significance of  \LamCNPi{} is $7.3\sigma$, which is calculated from the change of the likelihood values between  fits with and without the signal component  included, and accounting for the change in the number of degrees of freedom.

The $\LamC$ decay branching fractions ($\mathcal{B}$) are determined as 
\begin{equation} \label{eq:br}
  \mathcal{B}=\frac{N_{\mathrm{obs}}} {\sum_{ij} N_{ij}^{\mathrm{ST}}\cdot (\epsilon_{ij}^{\mathrm{DT}}/\epsilon_{ij}^{\mathrm{ST}}) },
\end{equation}
where the subscripts $i$ and $j$ represent the ST modes and the data samples at different c.m.~energies, respectively.
The parameters $N_{ij}^{\mathrm{ST}}$, $\epsilon_{ij}^{\mathrm{ST}}$ and $\epsilon_{ij}^{\mathrm{DT}}$ are the ST yields, ST efficiencies, and DT efficiencies, respectively.
The detection efficiencies $\epsilon_{ij}^{\mathrm{ST}}$ and $\epsilon_{ij}^{\mathrm{DT}}$ are estimated from  signal MC samples, where the key distributions of the ST modes have been reweighted to agree with those of the data.
The ST and DT efficiencies are summarized in Table~\ref{tab:yield-st-468} for data samples at a c.m.~energy of 4.682~GeV.
The detection efficiencies for the other data samples are summarized in the Supplemental Material~\cite{BESIII:Supplemental}.
The branching fractions are determined to be $\mathcal{B}(\LamCNPi)=(6.6\pm1.2\pm0.4)\times 10^{-4}$,
$\mathcal{B}(\LamCLamPi)=(1.31\pm0.08\pm0.05)\times 10^{-2}$, and 
$\mathcal{B}(\LamCSigPi)=(1.22\pm0.08\pm0.07)\times 10^{-2}$,
where the first uncertainties are statistical and the second systematic.

The systematic uncertainties for the branching fraction measurements comprise those associated with the ST yields, the $\pi^+$ tracking and PID efficiencies, the requirement of zero loose tracks, the determination of the DT signal yields, the decay branching fractions of $\Lambda$ and $\Sigma^0$ 
(for $\LamC\to \Lambda \pi^+$ and $\Sigma^0\pi^+$ only) and the statistical uncertainties from the MC samples.
The DT approach on which the measurement is based means that uncertainties associated with the ST selection efficiency cancel out.

The uncertainty in the ST yields is 0.5\%, which arises from the statistical uncertainty and a systematic component coming from the fit to the $\mBC$ distribution.
The uncertainties associated with the $\pi^+$ tracking and PID efficiencies are determined from studies of a control sample $J/\psi\to\pi^+\pi^-\pi^0$ decays, as explained in Ref.~\cite{Ablikim:2017systpi},
and are assigned to be 1.0\% and 2.0\%, respectively.
The uncertainty due to the zero loose-track requirement is 1.7\%, which is assigned from studies of a control sample of $\ee\to\LamC\ALamC$ with $\LamC\to p K^-\pi^+$ and the $\ALamC$ decaying into ten tag decay modes.
The uncertainties from the determination of the DT yields are 5.6\%, 2.5\% and 2.1\% for the decays $\LamCNPi$, $\Lambda \pi^+$, and $\Sigma^0 \pi^+$, respectively, including those from the fit range and 
the modeling of $q\bar{q}$ and $\LamC\ALamC$ backgrounds, which are estimated from varying the range and alternative polynomial descriptions for the $q\bar{q}$ and $\LamC\ALamC$ backgrounds, respectively. 
The uncertainties in the DT efficiencies due to the branching fractions of $\Lambda$ and $\Sigma^0$ are 1.4\% and 1.4\% for the decays $\LamCLamPi$ and $\Sigma^0 \pi^+$, respectively.
Uncertainties arising from the MC modeling are investigated by reweighting the MC distribution to data, and comparing the results obtained between the original and reweighted samples. 
The resultant uncertainties in the MC modeling are 0.8\% for $\LamCNPi$ and 3.8\% for $\LamCSigPi$, but negligible for $\LamCLamPi$.
The uncertainties associated with the finite size of the signal MC samples are 0.2\%. All other uncertainties are negligible.
Assuming that all the sources of bias are uncorrelated, the total uncertainties are then taken to be 
the quadratic sum of the individual values, which are 6.3\%, 4.0\%, and 5.4\% for 
the decays $\LamCNPi$, $\Lambda \pi^+$, and $\Sigma^0 \pi^+$, respectively.

In summary, the singly Cabibbo-suppressed decay \LamCNPi{} is observed  with a statistical significance of $7.3\sigma$ by using \ee{} collision data samples corresponding to a total integrated luminosity of 3.9~\ifb{} collected at  c.m. energies between 4.612 and 4.699~GeV with the BESIII detector.
 The branching fraction of \LamCNPi{} is measured to be $(6.6\pm1.2_{\rm stat}\pm0.4_{\rm syst})\times 10^{-4}$, this is a first-time measurement. 
Meanwhile, the branching fractions of the Cabibbo-favored decays \LamCLamPi{} and \LamCSigPi{} 
are measured to be $(1.31\pm0.08_{\rm stat}\pm0.05_{\rm syst})\times 10^{-2}$ and  $(1.22\pm0.08_{\rm stat}\pm0.07_{\rm syst})\times 10^{-2}$, respectively, which are consistent with previous BESIII results~\cite{Ablikim:2015flg}. 
The measured branching fraction of \LamCNPi{} is consistent with the prediction in Ref.~\cite{Geng:2018rse}, but twice as large as that in Ref.~\cite{Cheng:2018hwl}, implying that the nonfactorization contributions are overestimated. 
Taking the upper limit of the branching fraction of $\LamC\to p\pi^0$ from the Belle experiment, $\mathcal{B}(\LamC\to p\pi^0)<8.0 \times 10^{-5}$ at the 90\% confidence level~\cite{Li:2021uzo}, 
the ratio of branching fractions between \LamCNPi{} and $\LamC\to p\pi^0$ is calculated to be larger than 7.2 at the 90\% confidence level, which disagrees with the most predictions of  phenomenological models~\cite{Sharma:1996sc,  Lu:2016ogy, Uppal:1994pt, Cheng:2018hwl, Geng:2018plk, Geng:2018rse, HJZhao:2020}.
The results from this analysis provide an essential input for the phenomenological studies on the underlying dynamics of charmed baryon decays. 
In order to obtain an improved understanding it is desirable to perform improved studies of these decays, in particular concerning  the $\LamC\to p\pi^0$ branching fraction in the future~\cite{BESIII:whitepaper}.

The BESIII Collaboration thanks the staff of BEPCII, the IHEP computing center and the supercomputing center of the 
University of Science and Technology of China (USTC) for their strong support.
The authors are grateful to Hai-Yang Cheng, Fanrong Xu and Xianwei Kang for enlightening discussions. 
This work is supported in part by National Key R\&D Program of China under Contracts No. 2020YFA0406400, No. 2020YFA0406300; 
National Natural Science Foundation of China (NSFC) under Contracts No. 11335008, No. 11625523, No. 11635010, No. 11735014, No. 11822506, No. 11835012, No. 11935015, No. 11935016, No. 11935018, No. 11961141012, No. 12022510, No. 12025502, No. 12035009, No. 12035013, No. 12061131003, No. 12005311, No. 11805086, No. 11705192, No. 11950410506; 
the Fundamental Research Funds for the Central Universities, University of Science and Technology of China, Sun Yat-sen University, Lanzhou University, University of Chinese Academy of Sciences;
100 Talents Program of Sun Yat-sen University;
the Chinese Academy of Sciences (CAS) Large-Scale Scientific Facility Program; Joint Large-Scale Scientific Facility Funds of the NSFC and CAS under Contracts No. U1732263, No. U1832207, No. U1832103, No. U2032111; 
CAS Key Research Program of Frontier Sciences under Contract No. QYZDJ-SSW-SLH040; 100 Talents Program of CAS; China Postdoctoral Science Foundation under Contracts No. 2019M662152, No. 2020T130636;
INPAC and Shanghai Key Laboratory for Particle Physics and Cosmology; ERC under Contract No. 758462; 
European Union Horizon 2020 research and innovation programme under Marie Sklodowska-Curie Grant Agreement No. 894790; 
German Research Foundation DFG under Contracts No. 443159800, Collaborative Research Center CRC 1044, FOR 2359, FOR 2359, GRK 214; 
Istituto Nazionale di Fisica Nucleare, Italy; Ministry of Development of Turkey under Contract No. DPT2006K-120470; National Science and Technology fund; 
Olle Engkvist Foundation under Contract No. 200-0605; STFC (U.K.); 
The Knut and Alice Wallenberg Foundation (Sweden) under Contract No. 2016.0157; The Royal Society, U.K. under Contracts No. DH140054, No. DH160214; 
The Swedish Research Council; U. S. Department of Energy under Contracts No. DE-FG02-05ER41374, No.  DE-SC-0012069

\bibliography{bibitem}

\end{document}

\normalsize
\parskip=5pt plus 1pt minus 1pt

\title{ Supplementary materials of Observation of the Singly Cabbibo-Suppressed Decay \LamCNPi{} }

\input{author.tex}

\maketitle
\begin{table}[!htbp]
  \begin{center}
  \caption{The \dE{} requirement, ST yield,  the detection efficiency of ST and DT 
       for data sample with c.m.~energy at  4.612~GeV.
          The uncertainty in the ST yield is statistical only. }
  \renewcommand\arraystretch{1.2}
    \begin{tabular}{ l p{0.7cm}<{\raggedleft} @{,\, } p{0.7cm}<{\raggedright} p{1cm}<{\raggedleft} @{ $\pm$ } p{1cm}<{\raggedright} p{2.5cm}<{\centering} p{2.5cm}<{\centering} p{2.5cm}<{\centering} p{2.5cm}<{\centering} }
      \hline
      \hline
	    & \multicolumn{2}{c}{\dE{}(MeV)} & \multicolumn{2}{c}{$N_{i}^{\mathrm{ST}}$}  & $\epsilon_{i}^{\mathrm{ST}}$(\%) & $\epsilon_{i}^{\mathrm{DT}}(n\pi^+)$ (\%)  &  $\epsilon_{i}^{\mathrm{DT}}(\Lambda\pi^+)$ (\%) & $\epsilon_{i}^{\mathrm{DT}}(\Sigma^0\pi^+)$ (\%) \\
      \hline
	    $\apkpi$            & $(-34$  & $20)$    &   1,158 & 38  &  50.2 & 39.4  & 15.3  & 15.2\\
            $\apks$             &  $(-20$ & $20)$    &   241  & 16   &  53.2    &  43.0 & 16.6  & 16.2 \\
            $\apkpi\pi^0$     &  $(-30$ & $20)$      &  281 & 23  &  14.9   & 14.2  & 5.5  & 5.3 \\
            $\apks\pi^0$          & $(-30$ &  $20)$  &  109 & 13   &  17.2   &  15.7  & 6.0  & 5.8 \\
            $\apks\pi^+\pi^-$    & $(-20$ & $20)$    &  103 & 13    &  19.0  & 15.5 & 5.9  & 5.9 \\
            $\bar{\Lambda}\pi^-$      &  $(-20$ & $20)$     &  120 & 11   &  42.5  & 34.6 & 14.1 & 13.4 \\
            $\bar{\Lambda}\pi^-\pi^0$    & $(-30$ & $20)$ &  226 & 18  &  15.6   & 14.2 & 5.4 & 5.3 \\
            $\bar{\Lambda}\pi^+\pi^-\pi^-$ & $(-20$ & $20)$ &  128 & 13   &  12.7  & 10.5  & 4.1 & 4.1 \\
            $\bar{\Sigma}^0\pi^-$        & $(-20$ & $20)$  &  77 & 9   &  21.0   & 19.7   & 7.2 & 6.7 \\
            $\bar{\Sigma}^-\pi^+\pi^-$   & $(-30$ & $20)$ &  155 & 16   &  17.6  & 17.7  & 6.3 & 5.9 \\
      \hline\hline
    \end{tabular}%}
          \label{tab:yield-st-4612}
  \end{center}
\end{table}
\begin{table}[!htbp]
  \begin{center}
  \caption{The \dE{} requirement, ST yield,  the detection efficiency of ST and DT for data sample with c.m.~energy at  4.628~GeV.
          The uncertainty in the ST yield is statistical only. }
  \renewcommand\arraystretch{1.2}
    \begin{tabular}{ l p{0.7cm}<{\raggedleft} @{,\, } p{0.7cm}<{\raggedright} p{1cm}<{\raggedleft} @{ $\pm$ } p{1cm}<{\raggedright} p{2.5cm}<{\centering} p{2.5cm}<{\centering} p{2.5cm}<{\centering} p{2.5cm}<{\centering} }
      \hline
      \hline
	    & \multicolumn{2}{c}{\dE{}(MeV)} & \multicolumn{2}{c}{$N_{i}^{\mathrm{ST}}$}  & $\epsilon_{i}^{\mathrm{ST}}$(\%) & $\epsilon_{i}^{\mathrm{DT}}(n\pi^+)$ (\%)  &  $\epsilon_{i}^{\mathrm{DT}}(\Lambda\pi^+)$ (\%) & $\epsilon_{i}^{\mathrm{DT}}(\Sigma^0\pi^+)$ (\%) \\
      \hline
	    $\apkpi$            & $(-34$  & $20)$    &  5,911 & 87  &  49.5  & 39.0  & 14.9 & 14.7\\
            $\apks$             &  $(-20$ & $20)$   &  1,063 & 35  &  51.6  & 41.4  & 15.9 & 15.8\\
            $\apkpi\pi^0$     &  $(-30$ & $20)$        &  1,239 & 50  &  15.0   & 13.9  & 5.4 & 5.2 \\
            $\apks\pi^0$          & $(-30$ &  $20)$  &  460 & 29   &  17.0   & 15.4  & 5.8 & 5.7 \\
            $\apks\pi^+\pi^-$    & $(-20$ & $20)$    &  423 & 28   &   18.4   &  15.0  & 5.8 & 5.8 \\
            $\bar{\Lambda}\pi^-$      &  $(-20$ & $20)$     &  662 & 28   &   40.8   &  33.6 & 12.9 & 12.5 \\
            $\bar{\Lambda}\pi^-\pi^0$    & $(-30$ & $20)$ &  1,161 & 42  &   15.2   &  13.8  & 5.2 & 5.2 \\
            $\bar{\Lambda}\pi^+\pi^-\pi^-$ & $(-20$ & $20)$ & 512 & 29   &   12.5   &   10.1  & 3.9 & 3.9 \\
            $\bar{\Sigma}^0\pi^-$        & $(-20$ & $20)$  &  329 & 20   &   20.2   &  18.7  & 6.8  & 6.4 \\
            $\bar{\Sigma}^-\pi^+\pi^-$   & $(-30$ & $20)$ &  738 & 37   &   17.4   &  17.4  & 6.0 & 5.7\\
      \hline\hline
    \end{tabular}%}
          \label{tab:yield-st-4627}
  \end{center}
\end{table}
\begin{table}[!htbp]
  \begin{center}
  \caption{The \dE{} requirement, ST yield,  the detection efficiency of ST and DT for data sample with c.m.~energy at  4.641~GeV.
          The uncertainty in the ST yield is statistical only. }
  \renewcommand\arraystretch{1.2}
     \begin{tabular}{ l p{0.7cm}<{\raggedleft} @{,\, } p{0.7cm}<{\raggedright} p{1cm}<{\raggedleft} @{ $\pm$ } p{1cm}<{\raggedright} p{2.5cm}<{\centering} p{2.5cm}<{\centering} p{2.5cm}<{\centering} p{2.5cm}<{\centering} }
      \hline
      \hline
	    & \multicolumn{2}{c}{\dE{}(MeV)} & \multicolumn{2}{c}{$N_{i}^{\mathrm{ST}}$}  & $\epsilon_{i}^{\mathrm{ST}}$(\%) & $\epsilon_{i}^{\mathrm{DT}}(n\pi^+)$ (\%)  &  $\epsilon_{i}^{\mathrm{DT}}(\Lambda\pi^+)$ (\%) & $\epsilon_{i}^{\mathrm{DT}}(\Sigma^0\pi^+)$ (\%) \\
      \hline
	    $\apkpi$            & $(-34$  & $20)$    &  6,229 & 90  &  49.0  & 38.4 & 14.8 & 14.5 \\
            $\apks$             &  $(-20$ & $20)$   &  1,110 & 35  &  50.9  & 40.9  & 15.6 & 15.5 \\
            $\apkpi\pi^0$     &  $(-30$ & $20)$        &  1,307 & 52  &  14.8  & 13.9   & 5.3 & 5.2 \\
            $\apks\pi^0$          & $(-30$ &  $20)$  &  485 & 30   &  17.0  & 15.2  & 5.6 & 5.5 \\
            $\apks\pi^+\pi^-$    & $(-20$ & $20)$    &  455 & 28   &  18.4  & 14.9   & 5.8 & 5.8 \\
            $\bar{\Lambda}\pi^-$      &  $(-20$ & $20)$     &  691 & 28   &  40.4  & 33.0  & 12.5 & 12.4 \\
            $\bar{\Lambda}\pi^-\pi^0$    & $(-30$ & $20)$ &   1,328 & 45  &  15.3  & 13.7  & 5.1 & 5.0 \\
            $\bar{\Lambda}\pi^+\pi^-\pi^-$ & $(-20$ & $20)$ &   667 & 32   &  12.5  & 10.4  & 3.9 & 4.0 \\
            $\bar{\Sigma}^0\pi^-$        & $(-20$ & $20)$  &  345 & 21   &  20.5  & 18.4  & 6.7 & 6.2 \\
            $\bar{\Sigma}^-\pi^+\pi^-$   & $(-30$ & $20)$ &  812 & 38   &  17.1  & 17.1 & 5.9 & 5.5 \\
      \hline\hline
    \end{tabular}%}
          \label{tab:yield-st-4640}
  \end{center}
\end{table}
\begin{table}[!htbp]
  \begin{center}
  \caption{The \dE{} requirement, ST yield,  the detection efficiency of ST and DT for data sample with c.m.~energy at  4.661~GeV.
          The uncertainty in the ST yield is statistical only. }
  \renewcommand\arraystretch{1.2}
    \begin{tabular}{ l p{0.7cm}<{\raggedleft} @{,\, } p{0.7cm}<{\raggedright} p{1cm}<{\raggedleft} @{ $\pm$ } p{1cm}<{\raggedright} p{2.5cm}<{\centering} p{2.5cm}<{\centering} p{2.5cm}<{\centering} p{2.5cm}<{\centering} }
      \hline
      \hline
	    & \multicolumn{2}{c}{\dE{}(MeV)} & \multicolumn{2}{c}{$N_{i}^{\mathrm{ST}}$}  & $\epsilon_{i}^{\mathrm{ST}}$(\%) & $\epsilon_{i}^{\mathrm{DT}}(n\pi^+)$ (\%)  &  $\epsilon_{i}^{\mathrm{DT}}(\Lambda\pi^+)$ (\%) & $\epsilon_{i}^{\mathrm{DT}}(\Sigma^0\pi^+)$ (\%) \\
      \hline
	    $\apkpi$            & $(-34$  & $20)$    &  5,884 & 86  &  48.0   & 37.8 & 14.5 & 14.4 \\
            $\apks$             &  $(-20$ & $20)$   &  1,117 & 35  &  49.7   & 39.6  & 15.2  & 14.7 \\
            $\apkpi\pi^0$     &  $(-30$ & $20)$        &  1,349 & 54  &  14.6   & 13.5  & 5.1  & 5.0\\
            $\apks\pi^0$          & $(-30$ &  $20)$  &  479 & 30   &  16.5   & 14.8  & 5.5  & 5.6 \\
            $\apks\pi^+\pi^-$    & $(-20$ & $20)$    &  458 & 28   &  18.2   & 14.9  & 5.8 & 5.7 \\
            $\bar{\Lambda}\pi^-$      &  $(-20$ & $20)$     &  651 & 27   &  39.1   & 31.9 & 12.4  &  12.0 \\
            $\bar{\Lambda}\pi^-\pi^0$    & $(-30$ & $20)$ &   1,165 & 41  &  14.9   & 13.4  & 5.1 & 4.9 \\
            $\bar{\Lambda}\pi^+\pi^-\pi^-$ & $(-20$ & $20)$ &   624 & 30   &  12.7  & 10.2 & 3.9  & 3.9 \\
            $\bar{\Sigma}^0\pi^-$        & $(-20$ & $20)$  &  343 & 20   &  19.6   & 18.0  & 6.4 & 6.1\\
            $\bar{\Sigma}^-\pi^+\pi^-$   & $(-30$ & $20)$ &  751 & 36   & 16.7   & 16.7  & 5.8 & 5.4 \\
      \hline\hline
    \end{tabular}%}
          \label{tab:yield-st-4660}
  \end{center}
\end{table}
\begin{table}[!htbp]
  \begin{center}
  \caption{The \dE{} requirement, ST yield,  the detection efficiency of ST and DT  
           for data sample with c.m.~energy at  4.682~GeV.
          The uncertainty in the ST yield is statistical only. }
  \begin{tabular}{ l p{0.7cm}<{\raggedleft} @{,\, } p{0.7cm}<{\raggedright} p{1cm}<{\raggedleft} @{ $\pm$ } p{1cm}<{\raggedright} p{2.5cm}<{\centering} p{2.5cm}<{\centering} p{2.5cm}<{\centering} p{2.5cm}<{\centering} }
      \hline
      \hline
	    & \multicolumn{2}{c}{\dE{}(MeV)} & \multicolumn{2}{c}{$N_{i}^{\mathrm{ST}}$}  & $\epsilon_{i}^{\mathrm{ST}}$(\%) & $\epsilon_{i}^{\mathrm{DT}}(n\pi^+)$ (\%)  &  $\epsilon_{i}^{\mathrm{DT}}(\Lambda\pi^+)$ (\%) & $\epsilon_{i}^{\mathrm{DT}}(\Sigma^0\pi^+)$ (\%) \\
      \hline
	    $\apkpi$            & $(-34$  & $20)$    &  17,415  & 145  & 47.3  &  37.0 & 14.2 & 14.1 \\
            $\apks$             &  $(-20$ & $20)$   &  3,353 & 61  &  48.1   & 38.8  & 14.8 & 14.5 \\
            $\apkpi\pi^0$     &  $(-30$ & $20)$        &  4,005 & 95  & 14.5  & 13.4 & 5.1 & 5.0 \\
            $\apks\pi^0$          & $(-30$ &  $20)$  &  1,454 & 52    &  16.5  & 14.4 & 5.5 & 5.4 \\
            $\apks\pi^+\pi^-$    & $(-20$ & $20)$    &  1,261 & 49   &  17.7  & 14.8 & 5.5 & 5.6 \\
            $\bar{\Lambda}\pi^-$      &  $(-20$ & $20)$     &  2,012 & 47   &  37.8  & 31.0 & 12.0 & 11.6 \\
            $\bar{\Lambda}\pi^-\pi^0$    & $(-30$ & $20)$ &  3,576 & 71  &  14.6  & 12.9 & 4.9 & 4.8 \\
            $\bar{\Lambda}\pi^-\pi^+\pi^-$ & $(-20$ & $20)$ &  1,818 & 52   &  12.3  & 10.3 & 4.0 & 3.8 \\
            $\bar{\Sigma}^0\pi^-$        & $(-20$ & $20)$  &  1,047 & 34   &  19.3  & 17.4 & 6.1 & 5.9 \\
            $\bar{\Sigma}^-\pi^+\pi^-$   & $(-30$ & $20)$ &  2,275 &  63  &  16.2  & 16.1 & 5.6 & 5.2 \\
      \hline\hline
    \end{tabular}%}
          \label{tab:yield-st-468-2}
  \end{center}
\end{table}
\begin{table}[!htbp]
  \begin{center}
  \caption{The \dE{} requirement, ST yield,  the detection efficiency of ST and DT for data sample with c.m.~energy at  4.699~GeV.
          The uncertainty in the ST yield is statistical only. }
  \renewcommand\arraystretch{1.2}
    \begin{tabular}{ l p{0.7cm}<{\raggedleft} @{,\, } p{0.7cm}<{\raggedright} p{1cm}<{\raggedleft} @{ $\pm$ } p{1cm}<{\raggedright} p{2.5cm}<{\centering} p{2.5cm}<{\centering} p{2.5cm}<{\centering} p{2.5cm}<{\centering} }
      \hline
      \hline
	    & \multicolumn{2}{c}{\dE{}(MeV)} & \multicolumn{2}{c}{$N_{i}^{\mathrm{ST}}$}  & $\epsilon_{i}^{\mathrm{ST}}$(\%) & $\epsilon_{i}^{\mathrm{DT}}(n\pi^+)$ (\%)  &  $\epsilon_{i}^{\mathrm{DT}}(\Lambda\pi^+)$ (\%) & $\epsilon_{i}^{\mathrm{DT}}(\Sigma^0\pi^+)$ (\%) \\
      \hline
	    $\apkpi$            & $(-34$  & $20)$    &  5,156 & 80  &  46.4  & 36.4 & 13.9 & 13.8\\
            $\apks$             &  $(-20$ & $20)$   &  964  & 33  &  47.3  & 37.5  & 14.4 & 14.1 \\
            $\apkpi\pi^0$     &  $(-30$ & $20)$        &  1,116 & 51  &  14.2  & 13.0   & 5.0  & 4.9\\
            $\apks\pi^0$          & $(-30$ &  $20)$  &  386 & 26   &  16.1  & 14.1  & 5.3  & 5.2 \\
            $\apks\pi^+\pi^-$    & $(-20$ & $20)$    &  417 & 27  &  17.6  & 14.5  & 5.6  & 5.5 \\
            $\bar{\Lambda}\pi^-$      &  $(-20$ & $20)$     &  519 & 24   &  37.1  & 30.3  & 11.6  & 11.7  \\
            $\bar{\Lambda}\pi^-\pi^0$    & $(-30$ & $20)$ &   1,045 & 39 &  14.2  & 12.7  & 4.8  & 4.7 \\
            $\bar{\Lambda}\pi^+\pi^-\pi^-$ & $(-20$ & $20)$ &   548 & 28   &  12.5  & 10.2 & 3.9  & 4.0 \\
            $\bar{\Sigma}^0\pi^-$        & $(-20$ & $20)$  &  283 & 18   &  18.8  & 16.9  & 6.3 & 5.9 \\
            $\bar{\Sigma}^-\pi^+\pi^-$   & $(-30$ & $20)$ &  699 & 35   &  16.1  & 15.8  & 5.5 & 5.1 \\
      \hline\hline
    \end{tabular}%}
          \label{tab:yield-st-4700}
  \end{center}
\end{table}
\begin{figure}[!htp]
    \begin{center}
        \includegraphics[width=0.42\textwidth]{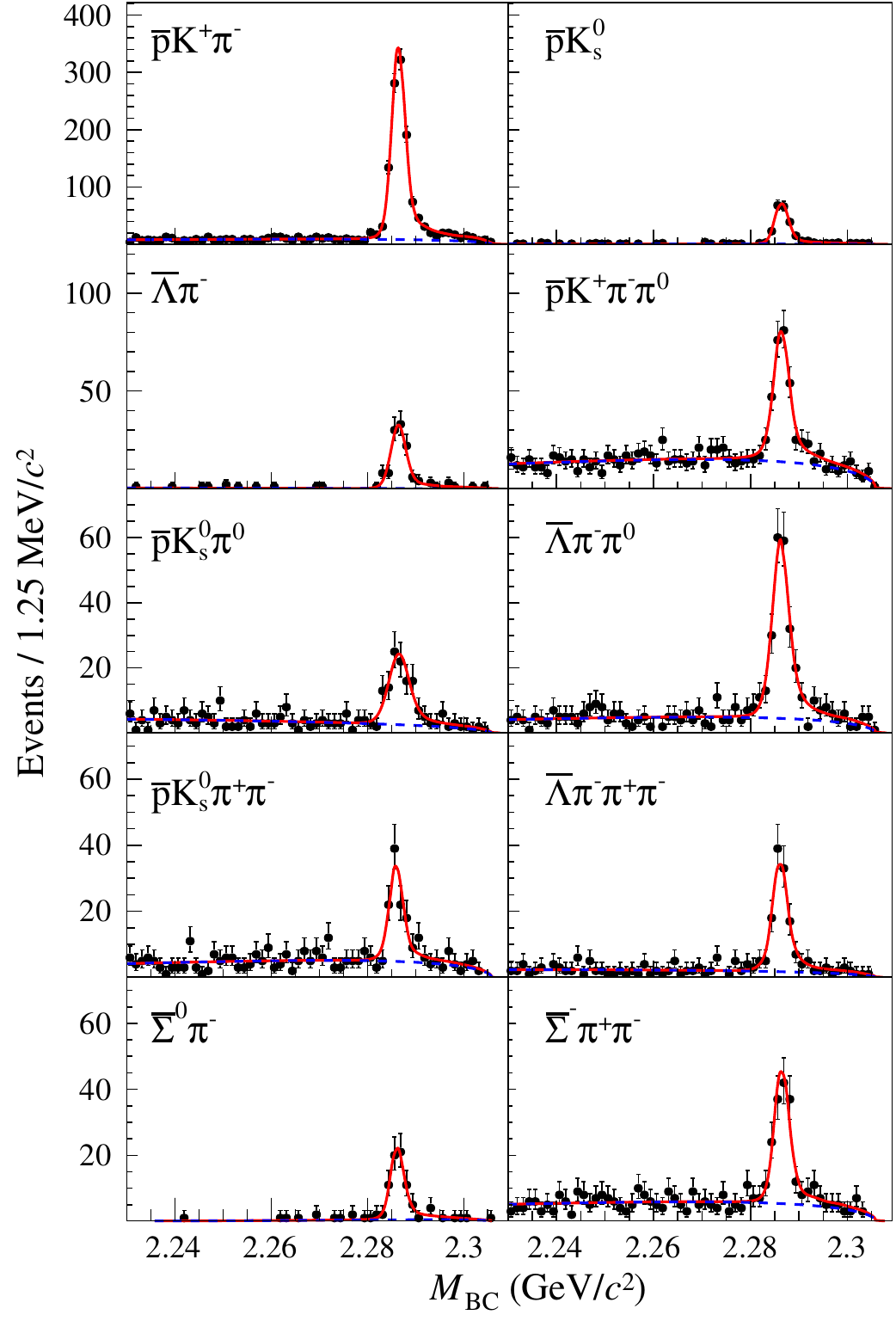}
    \end{center}
    \caption{
      The \mBC{} distributions and the fit curves of the individual ST mode for data sample with c.m.~energy at 4.612~GeV. }
     \label{fig:single-tag-4612}
\end{figure}
\begin{figure}[!htp]
    \begin{center}
        \includegraphics[width=0.42\textwidth]{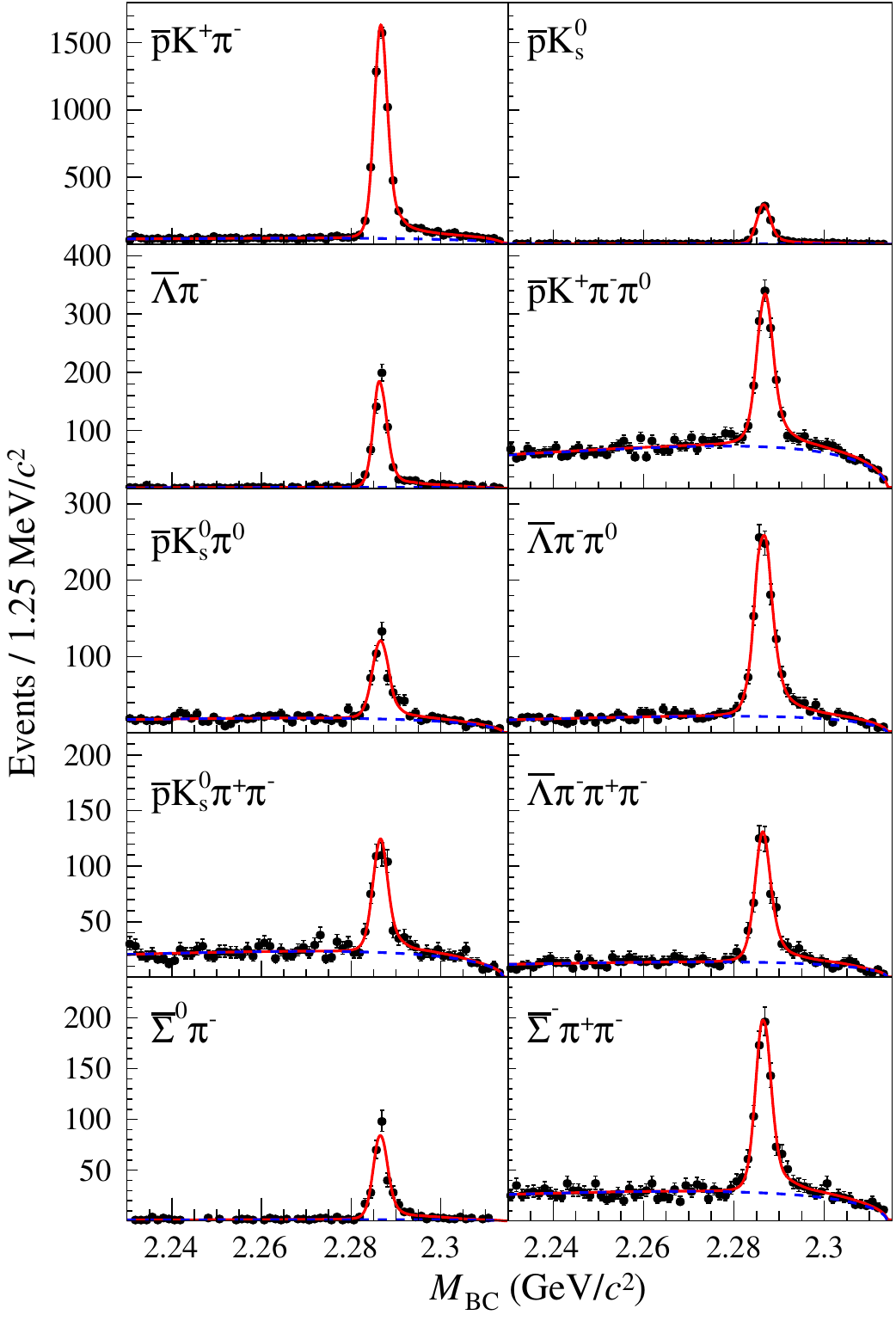}
    \end{center}
    \caption{
      The \mBC{} distributions and the fit curves of the individual ST mode for data sample with c.m.~energy at 4.628~GeV. }
     \label{fig:single-tag-462}
\end{figure}
\begin{figure}[!htp]
    \begin{center}
        \includegraphics[width=0.42\textwidth]{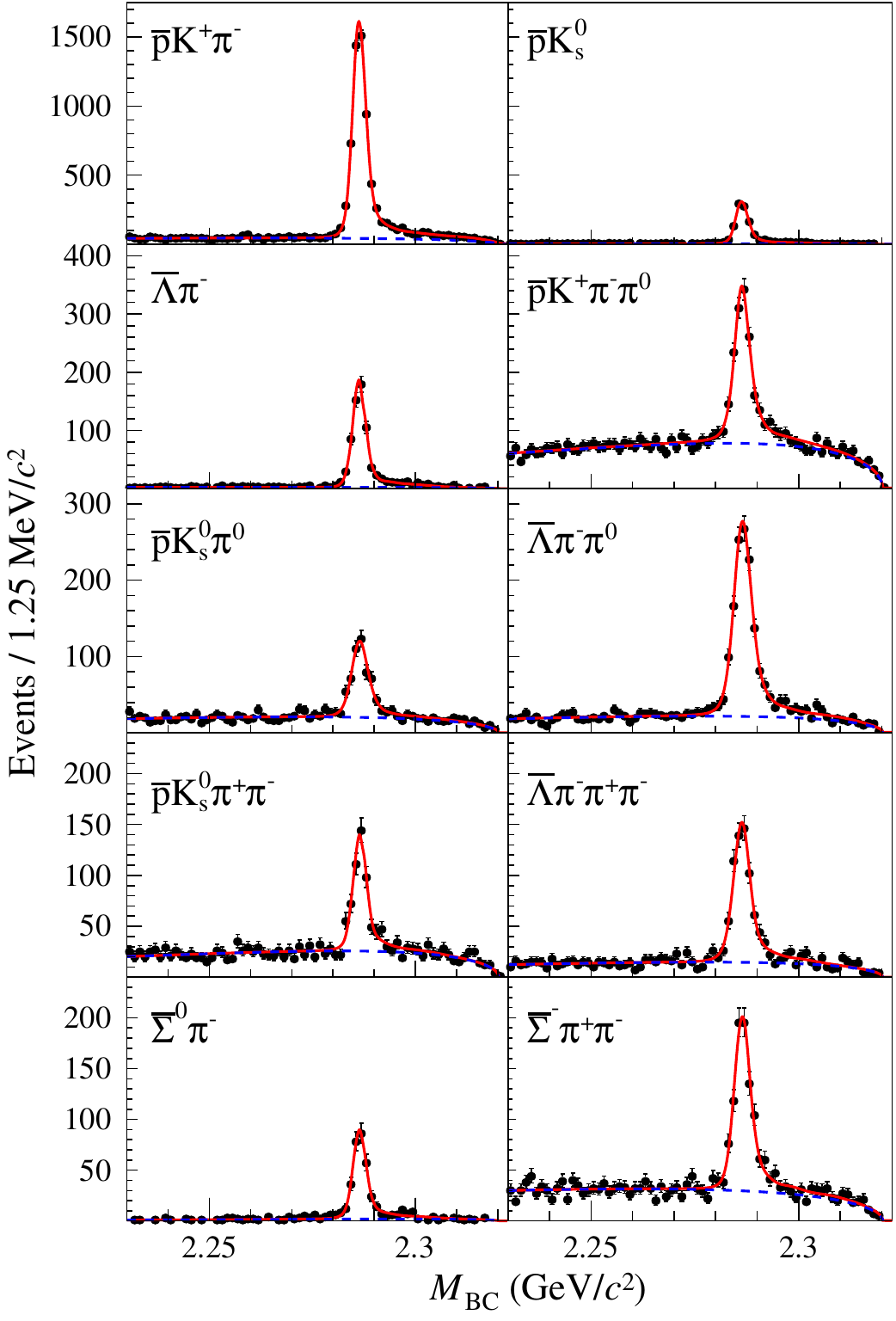}
    \end{center}
    \caption{
      The \mBC{} distributions and the fit curves of the individual ST mode for data sample with c.m.~energy at 4.641~GeV. }
     \label{fig:single-tag-464}
\end{figure}
\begin{figure}[!htp]
    \begin{center}
        \includegraphics[width=0.42\textwidth]{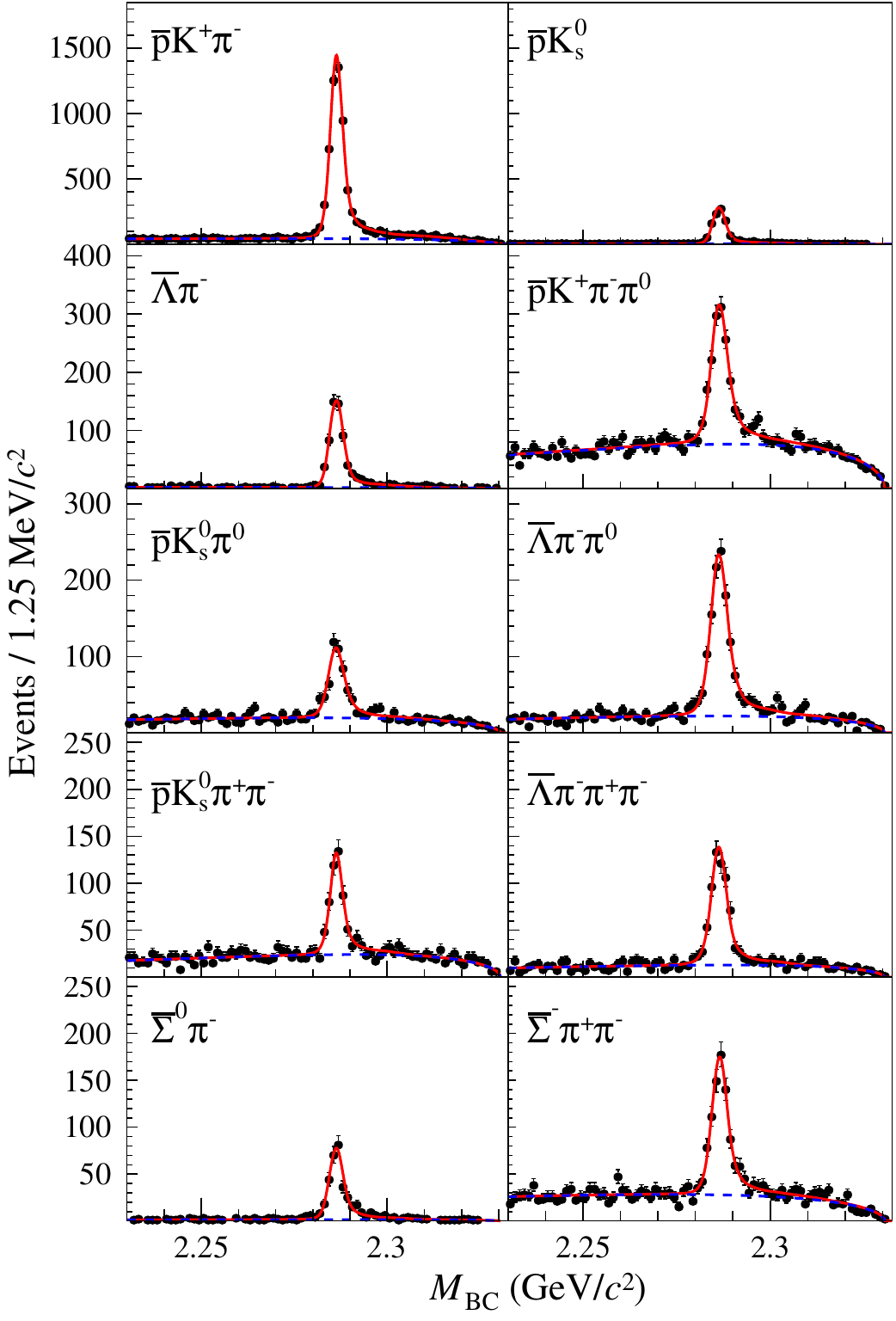}
    \end{center}
    \caption{
      The \mBC{} distributions and the fit curves of the individual ST mode for data sample with c.m.~energy at 4.661~GeV. }
     \label{fig:single-tag-466}
\end{figure}
\begin{figure}[!htp]
    \begin{center}
        \includegraphics[width=0.42\textwidth]{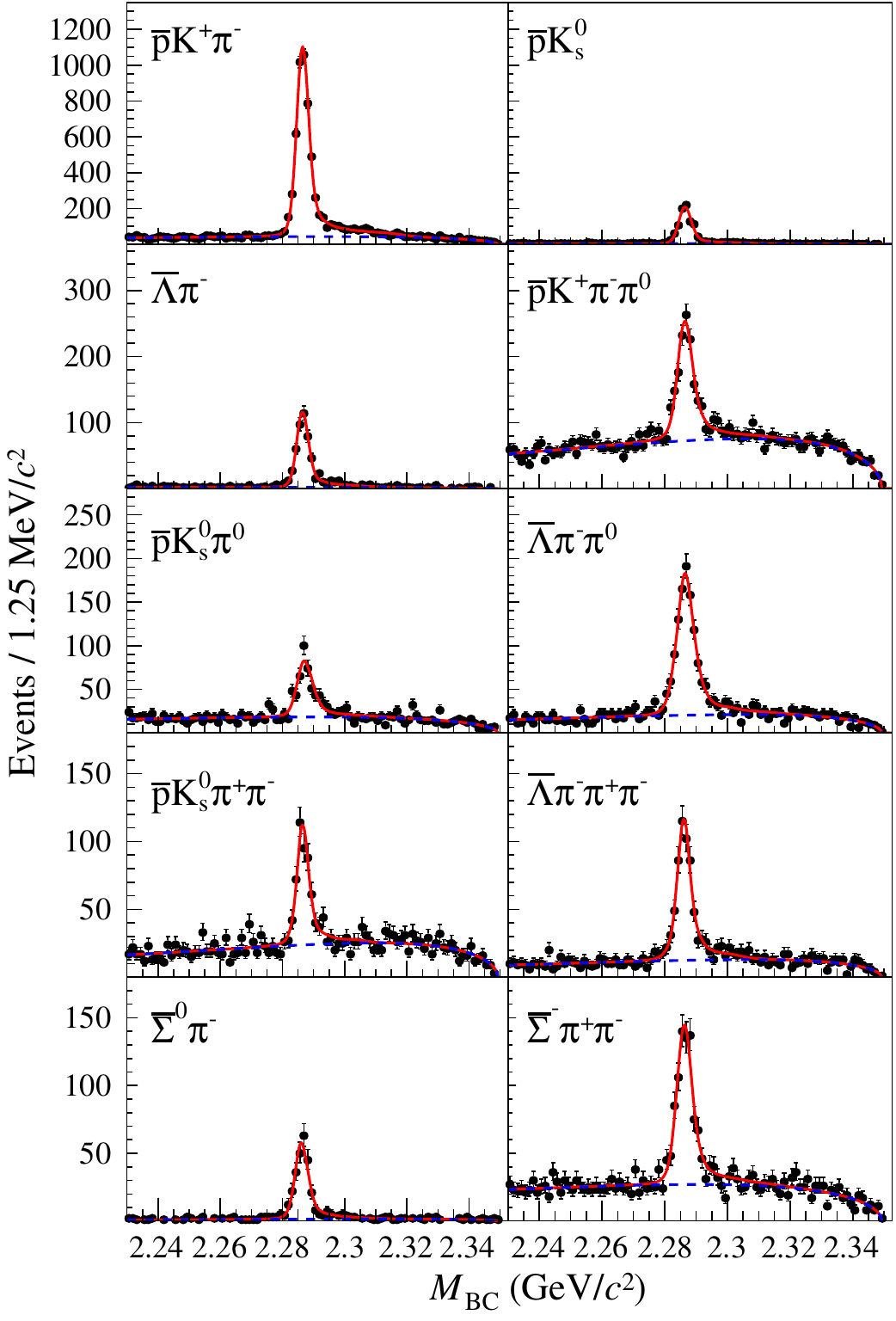}
    \end{center}
    \caption{
      The \mBC{} distributions and the fit curves of the individual ST mode for data sample with c.m.~energy at 4.699~GeV. }
     \label{fig:single-tag-470}
\end{figure}

%% file: author.tex
\author{
	\begin{small}
		\begin{center}
			M.~Ablikim$^{1}$, M.~N.~Achasov$^{10,b}$, P.~Adlarson$^{68}$, S. ~Ahmed$^{14}$, M.~Albrecht$^{4}$, R.~Aliberti$^{28}$, A.~Amoroso$^{67a,67c}$, M.~R.~An$^{32}$, Q.~An$^{64,50}$, X.~H.~Bai$^{58}$
% ...
			\\
			\vspace{0.2cm}
			(BESIII Collaboration)\\
			\vspace{0.2cm} {\it
				$^{1}$ Institute of High Energy Physics, Beijing 100049, People's Republic of China\\
% ...
$^{4}$ Bochum Ruhr-University, D-44780 Bochum, Germany\\
% ...
$^{10}$ G.I. Budker Institute of Nuclear Physics SB RAS (BINP), Novosibirsk 630090, Russia\\
% ...
$^{14}$ Helmholtz Institute Mainz, Staudinger Weg 18, D-55099 Mainz, Germany\\
% ...
$^{28}$ Johannes Gutenberg University of Mainz, Johann-Joachim-Becher-Weg 45, D-55099 Mainz, Germany\\
% ...
$^{32}$ Liaoning Normal University, Dalian 116029, People's Republic of China\\
% ...
$^{50}$ State Key Laboratory of Particle Detection and Electronics, Beijing 100049, Hefei 230026, People's Republic of China\\
% ...
$^{58}$ University of Jinan, Jinan 250022, People's Republic of China\\
% ...
$^{64}$ University of Science and Technology of China, Hefei 230026, People's Republic of China\\
% ...
$^{67}$ University of Turin and INFN, (A)University of Turin, I-10125, Turin, Italy; (B)University of Eastern Piedmont, I-15121, Alessandria, Italy; (C)INFN, I-10125, Turin, Italy\\
				$^{68}$ Uppsala University, Box 516, SE-75120 Uppsala, Sweden\\
% ...
$^{b}$ Also at the Novosibirsk State University, Novosibirsk, 630090, Russia\\
% ...
		}\end{center}	
		\vspace{0.4cm}
	\end{small}
}